# Storing events to retell them

Jean-Louis Dessalles
*ParisTech, Ecole Nationale Superieure des Telecommunications, F-75013 Paris, France.*
**dessalles@enst.fr   www.enst.fr/~jld**

**Abstract:** Episodic memory is certainly a unique endowment, but its primary purpose is something other than to provide raw material for creative synthesis of future scenarios. Remembered episodes are exactly those that are worth telling. The function of episodic memory, in my view, is to accumulate stories that are relevant to recount in conversation.

As the authors of the target article suggest, episodic memory (EM) can be seen as a "plug-in" device added to a standard vertebrate brain, and quite an expensive one, as much of our cortical mass seems devoted to it. Suddendorf & Corballis (S&C) are certainly right to say that it demands an evolutionary explanation. Their suggestion is that EM serves only the practical purpose of providing raw material for future planning. In a similar attempt to provide an evolutionary account for EM, Brown and Kulik (1977) highlighted the benefit of storing unexpected and highly emotional events, as "a marked departure from the ordinary in a consequential domain would leave [the individual] unprepared to respond adequately and endanger his survival" (p. 97). The problem with such accounts is that EM is badly designed for its alleged function.

Does an optimal use of storage capacity leave room for the memory of instantiated episodes? In machine learning, rote learning is an inefficient strategy. The purpose of any learning task is to make generalization possible. A good way to perform induction is to aggregate experience into structures such as prototypes or clusters. Storing particular instances (e.g., cluster centers) generally makes sense if they are statistically representative. This function is implemented in living beings through semantic and procedural memory.

To delineate categories, it may be useful to remember borderline instances, as with support vectors (Cornuejols & Miclet 2002). Also, in certain applications in which data are scarce and non-homogeneous, storing actual encountered examples, regardless of their representativeness, may be a viable strategy. The Case-Based Reasoning technique (Kolodner 1993) aims at solving new problems by matching them with memorized known examples. Superficially, episodic memory could be understood as a biological implementation of these principles, but its actual form does not match up to the assignment.

Episodic memory is highly selective. It retains a tiny fraction of all our daily experiences. One may come across dozens of people each day and remember only a few encounters per month. Selected episodes are, however, retained with a significant amount of detail, including what S&C call the www criterion. From an efficiency perspective, details such as the precise location in space and time, the weather conditions, the persons present, the words exchanged, and so forth, are most often irrelevant and yet are almost systematically remembered, even in the long term in cases when emotion is high (Brown

& Kulik 1977). From a computational perspective, not only do such details represent a waste of storage, they also hinder and mislead retrieval matching.

An alternative view is that EM is an outgrowth of the language faculty (Dessalles 2006). It is not fortuitous that *memorized episodes are exactly those which are narratable.* People spend one fifth of their waking time in spontaneous conversation (Dunbar 1998), and a significant share of this time is devoted to reporting past events (Tannen 1984, p. 99; Eggins & Slade 1997, p. 265). Interlocutors draw from their memory relevant episodes that they can relate to the current conversational topic and they systematically try to recount them. However, only a tiny fraction of past experiences may be recounted in this way. One crucial requirement is that reported stories must appear *unexpected* (Dessalles, in press).

The requirement of unexpectedness provides also a good prediction of the kind of episodes that are preferentially stored in memory. To appear unexpected, a situation must be *less complex* (i.e., more easily describable) than expected (Dessalles, in press). Witnessing a six-legged cow makes both a memorable event and a good story to tell, just because this cow, thanks to its unique peculiarity, requires a minimal description to be distinguished from all others. If, as we claim, the primary purpose of storing episodes is to offer material for future recounting, then systematically remembering details such as time and space location makes perfect sense. If the six-legged animal lives in the vicinity, interest is raised. Not specifying the location would leave the listener with the idea that that location requires a lengthy description, and interest drops down. By computing complexity differences, one can derive the way interest varies according to location and time, and according to various factors such as the persons involved (Dessalles, in press). For instance, the interest of coincidentally encountering someone increases with the complexity (*e.g.* remoteness) of the place and the simplicity (*e.g.* celebrity or closeness) of that person. It is thus crucial, when memorizing an episode, to remember every detail that may affect the cognitive complexity of the situation.

It may seem surprising that the expensive resources devoted to EM serve such a futile purpose as everyday chatter. This is only because one fails to see that casual conversation is an arena where much of our social existence is decided (Dessalles 2007). Eliciting interest through conversational stories is a high-stakes game. Boring participants are rapidly ignored and may lose their friends. When it comes to establishing solidarity bonds, individuals prefer those who successfully demonstrate their informational capacity and their experience with unexpected events. In our species, those who know first or who can draw highly relevant events from their past experiences make potentially good allies. Natural selection favored not only this preference, but also the narrative skills that allow any of us to display these qualities (Dessalles 2007). Episodic memory, in this context, is a crucial tool that enables us to produce the most relevant story at the right time. It has been tailored for this purpose, as demonstrated by the fact that the factors that favor memorization, such as unexpectedness and atypicality (Shapiro & Fox 2002; Stangor & McMillan 1992; Woll & Graesser 1982), are exactly the factors which increase tellability.

This account explains why remembered episodes are communicated, instead of remaining private; why they remain coherent in memory (instead of being dismantled for creative synthesis of future scenarios); why they systematically involve various details and precision; why we keep on memorizing episodes throughout our entire life; and why even slight failures in episodic memory (as those that occur with aging or in certain pathologies) have dramatic influence on social relations. It also explains the uniqueness of EM, which was not selected for increasing planning efficiency, but as a tool in support of language performance.

## References

Brown, R. & Kulik J. (1977) Flashbulb memories. *Cognition* 5:73–99.

Cornuejols, A. & Miclet, L. (2002) *Apprentissage artificiel – Concepts et algorithmes* [Machine Learning: Concepts and algorithms]. Eyrolles.

Dessalles, J.-L. (2006) Human language in the light of evolution. In: *JEP 2006: Actes des XXVIes journées d'étude sur la parole, AFCP – IRISA – ISCA*, pp. 17–23. Dinard. Available at: http://www.enst.fr/~jld/papiers/pap.evol/Dessalles_06072301.pdf

Dessalles, J.-L. (2007) *Why we talk – The evolutionary origins of language.* Oxford University Press. Available at: http://www.enst.fr/~jld/WWT/

Dessalles, J.-L. (in press) Complexité cognitive appliquée à la modélisation de l'intérêt narratif [Cognitive complexity as a determining factor of narrative interest]. *Intellectica* 45.

Dunbar, R. I. M. (1998) Theory of mind and the evolution of language. In: *Approaches to the evolution of language: Social and cognitive bases,* ed. J. R. Hurford, M. Studdert-Kennedy & C. Knight, pp. 92–110. Cambridge University Press.

Eggins, S. & Slade, D. (1997) *Analysing casual conversation.* Equinox.

Kolodner, J. (1993) *Case-based reasoning.* Morgan Kaufmann.

Shapiro, M. A. & Fox, J. R. (2002) The role of typical and atypical events in story memory. *Human Communication Research* 28(1):109–35.

Stangor, C. & McMillan, D. (1992) Memory for expectancy-congruent and expectancy-incongruent information: a review of the social and social developmental literatures. *Psychological Bulletin* 111(1):42–61.

Tannen, D. (1984) *Conversational style – Analyzing talk among friends.* Ablex.

Woll, S. & Graesser, A. (1982) Memory discrimination for information typical and atypical of person schemata. *Social Cognition* 1:287–310.